\documentstyle[aps]{revtex}                                 

\begin{document}


\newcommand{\gm}[1]{\gamma^{#1}}
\newcommand{\Gm}[1]{\Gamma^{#1}}
\newcommand{\vb}[2]{e^{#1}_{#2}}
\newcommand{\vbi}[2]{e_{#1}^{#2}}
\newcommand{\pa}{\partial}
\newcommand{\SUSY}{supersymmetry\ }
\newcommand{\SUSYs}{supersymmetries\ }
\newcommand{\hF}{\hat{F}}
\newcommand{\arccot}{\cot^{-1}}

 
\newcommand{\beq}{\begin{equation}}
\newcommand{\beql}[1]{\begin{equation}\label{eq:#1}}
\newcommand{\eeq}{\end{equation}}
\newcommand{\beqn}{\begin{eqnarray}}
\newcommand{\eeqn}{\end{eqnarray}}
\newcommand{\eq}[1]{(\ref{eq:#1})}


\newcommand{\npb}[3]{Nucl. Phys. {\bf B#1} (#2) #3}
\newcommand{\plb}[3]{Phys. Lett. {\bf #1B} (#2) #3}
\renewcommand{\prd}[3]{Phys. Rev. {\bf D#1} (#2) #3}
\renewcommand{\prl}[3]{Phys. Rev. Lett. {\bf #1} (#2) #3}
\newcommand{\mpl}[3]{Mod. Phys. Lett. {\bf A#1} (#2) #3}
\newcommand{\hepth}[1]{hep-th/#1}


\title{D-Branes, Moduli and Supersymmetry}
\author{ 	Vijay Balasubramanian\thanks{vijayb@puhep1.princeton.edu}}
\address{	Joseph Henry Laboratories, Princeton University, 
			Princeton, NJ 08544}
\and	
\author{ 	Robert G. Leigh\thanks{rgleigh@uiuc.edu}}   
\address{	Department of Physics, 
			University of Illinois at Urbana-Champaign, 
			Urbana, IL 61801} 
\maketitle

\begin{abstract}
We study toroidal compactifications of Type II string theory with
D-branes and nontrivial antisymmetric tensor moduli and show that
turning on these fields modifies the supersymmetry projections imposed
by D-branes.  These modifications are seen to be necessary for the
consistency of T-duality. We also show the existence of unusual BPS
configurations of branes at angles that are supersymmetric because of
conspiracies between moduli fields.  Analysis of the problem from the
point of view of the effective field theory of massless modes shows that
the presence of a 2-form background must modify the realization of
supersymmetry on the brane. In particular, the appropriate supersymmetry
variation of the physical gaugino vanishes in any constant field
strength background. These considerations are relevant for the
$E_{7(7)}$-symmetric counting of states of 4-dimensional black holes in
Type II string theory compactified on $T^6$.
\end{abstract}

\section{Introduction}

D-branes have played an important role in recent advances in string
theory.  The realization by Polchinski~\cite{joeRR} that D-branes are
the carriers of Ramond-Ramond (RR) charges, coupled with detailed
analyses of the interactions of D-branes with string
modes~\cite{dlp,rgl}, has contributed greatly to the current
understanding of non-perturbative aspects of string theory.  D-branes
are BPS states, and so break half of the available supersymmetry and
configurations of more than one D-brane can break supersymmetry further.
Configurations consisting of various types of D-branes intersecting at
right angles have been studied by many authors. Indeed, this has been
important in the analysis of black hole entropy using D-brane
states~\cite{vafastrom,bhetc}.  The corresponding brane configurations
have also been studied in M-theory~\cite{ggetc}. Recently it has been
realized~\cite{bdl} that more general brane configurations can preserve
some supersymmetry - branes may intersect at angles, given some simple
restrictions on those angles.

In this note we extend the work of Ref.~\cite{bdl} by discussing
toroidal compactifications of D-branes in Type II string theory with
moduli.  In Sec.~\ref{sec:setup} we present the general formulation of
D-brane compactifications on bent tori with NS 2-form backgrounds, and
show that turning on $B_{\mu\nu}$ modifies the \SUSY projections imposed
by the presence of the D-branes.  A non-vanishing $B_{\mu\nu}$ or world-volume
gauge field on a $p$-brane also induces the charges of smaller branes via
world-volume Chern-Simons couplings. In Sec.~\ref{sec:review} we review the
general solution of branes at angles given in~\cite{bdl} and develop
techniques that will be used in subsequent sections.  In
Sec.~\ref{sec:examples} we apply T-duality to BPS configurations of
branes at angles and show that the general formulation of
Sec.~\ref{sec:setup} is needed to correctly account for BPS saturation
after the duality transformation.  Including the effects of the 2-form
moduli leads to some surprises.  For example, as discussed in
Sec.~\ref{sec:examples}, a 1-brane orthogonal to a 5-brane can be
supersymmetric if a suitable background B field (or gauge field on the
5-brane) is turned on.  In fact, after compactifying on a 6-torus there
are spaces of $N=1, \, d=4$ supersymmetric configurations of
1-branes at angles to 5-branes.  In the limit that the 1-brane is
embedded in the 5-brane, these become $N=2, d=4$ configurations.

In Sec.~\ref{sec:gauge} we examine these BPS configurations from the
point of view of the effective field theory on the brane and conclude
that the non-vanishing 2-form background must modify the realization of
world-volume supersymmetry.  In particular, the \SUSY variation of the
physical gaugino field must be modified in the presence of a background
field strength.  Working in light cone gauge, we derive the modified
variation and show that any constant 2-form background (not necessarily
self-dual or anti-self-dual) preserves world-brane supersymmetry.   This
cannot be understood by keeping the leading (Yang-Mills) terms in the
world-volume effective theory and suggests that the non-linear structure
of the Dirac-Born-Infeld D-brane effective action must come into play in
an interesting way.

Finally, in Sec.~\ref{sec:classic} we discuss briefly the relation of
the D-brane configurations uncovered here to classical p-brane solutions
of the Type II supergravity theories.  In particular, the general black
hole solutions of these theories compactified on a six-torus contain
branes compactified at angles.  Therefore, the considerations of this
paper are relevant for the $E_{7(7)}$-symmetric~\cite{ch} counting of
states of 4-dimensional black holes in Type II string
theory~\cite{vbrl}.

\section{General Formulation}
\label{sec:setup}
In this section we will derive the \SUSY projections imposed by the
presence of D-branes compactified on tori with non-vanishing NS 2-form
moduli.  Since there is a local symmetry connecting the world-volume 
field-strength $F$ with the (pullback of the) NS-NS 2-form $B$, 
it is $\hF=F+B$ that will appear.
We begin with a single 9-brane in the absence of any
background fields. The supersymmetry condition is (in the boundary
state formalism):
\begin{equation}
(Q^A+\Omega_9(\Gamma)\ \tilde Q^A) |B\rangle =0
\label{eq:ninezero}
\end{equation}
where $Q^A$ is a spacetime supersymmetry generator
 and is just the zero-momentum fermion vertex
operator taken in some fixed ghost picture.
 The factor $\Omega_9(\Gamma)\equiv\pm\Gamma_{11}$ 
in~\eq{ninezero} refers to the orientation of
the brane.\footnote{We take $\Gamma_{11}\tilde Q=+\tilde Q$.}
In a constant background gauge field, Ref.~\cite{clny}
gives the corresponding relation:
\begin{equation}
(Q+\Omega_9(\Gamma)\  M(\hF)\;\tilde Q)^A |B\rangle =0.
\label{nineF}
\end{equation}
(Also see~\cite{miao} in a D-brane context.)
Turning on $\hF$ produces a relative rotation of left and right moving
vectors by the matrix $R=\left( {1-\hF\over 1+\hF}\right)$, and $M(\hF)$ is
just the spinorial representation of this rotation.  For example, if $\hF$
is $2\times2$:
\begin{equation}
\hF=\pmatrix{0&a\cr -a&0} \: \: \: \: \:
a\equiv\tan\beta/2
\end{equation}
then $R$ is just a rotation by the angle $\beta$.  $M(F)$ is given by:
\begin{equation}
M(\hF)=\det (1+\hF)^{-1/2}\; \AE(-1/2\; \hF_{\mu\nu}\gamma^\mu\gamma^\nu)
\label{M}
\end{equation}
where $\AE$ is the exponential function defined through its power
series expansion, with the product of $\gamma$-matrices in each term
antisymmetrized.

Now consider the case of a general $p$D-brane with both metric and
antisymmetric tensor backgrounds.  Let $\{ \vb{i}{\mu} \}$ and $\{
f^j_\mu \}$ be vielbeins spanning the spaces tangent and normal to the
D-brane and let $\Gm{i} = \vb{i}{\mu}\gm{\mu}$ and $\tilde{\Gamma}^j =
f^j_\mu \gamma^\mu$.  We start with eq.~\eq{ninezero} for a 9-brane
and construct $p$D-branes by T-dualizing $9-p$ of the dimensions.
T-duality changes the sign of the right-moving coordinates and
worldsheet supersymmetry requires a similar change in the worldsheet
fermions. The spacetime supersymmetry generators are built out of
zero-momentum Ramond vertex operators and so we expect that the
right-moving supersymmetry generator picks up a factor of
$\tilde{\Gamma}^{i} \Gamma^{11}$ for each T-dualized
dimension~\cite{Dnotes}.  This gives the supersymmetry relation
$\left[Q \pm \prod_j \left( \tilde{\Gamma}_j \Gamma_{11} \right)
\tilde{Q}\right]^A |B\rangle = 0 $.  When an $\hF$ background on the
D-brane is turned on, open strings sense the background through their
endpoints, causing a Lorentz-rotation of right-moving fields relative
to left-moving fields.  This further modifies the relation to $\left[Q
\pm \prod_j \left(\tilde{\Gamma}_j \Gamma_{11} \right) M(\hat{F})
\tilde{Q}\right]^A |B\rangle = 0 $.  Commuting the factors of
$\Gamma_{11}$ through the other gamma matrices we arrive at the
supersymmetry condition:
\begin{equation}
(Q+\Omega_p(\Gamma)
 M(\hF)\;\tilde Q)^A |B\rangle =0.
\label{basic}
\end{equation}
where $\Omega$ is the volume form of the brane (normalized to unity)
and 
\beq
\Omega_p(\Gamma)=\Omega_{i_1\ldots i_{p+1}}\Gm{i_1}\ldots\Gm{i_{p+1}}.
\eeq
The sign is fixed by comparing to the 9-brane equation~\eq{ninezero}.

We may also check the form of eq.~(\ref{basic}), at least for
branes of Type IIB, by taking a limit of eq.~(\ref{nineF}). The
boundary condition for open strings in a background field takes
the form:
\beql{bdycon}
\pa_n X_i + \left( g^{-1}\hF\right)_{ij} \pa_\tau X^j =0
\eeq
Thus, if we take $\hF$ from zero to infinity, we will convert two 
boundary conditions from Neumann to Dirichlet. In fact, there are
many directions in which to go to infinity, and these correspond
via T-duality to D-branes at angles.  We will explore this in detail
in a later  section, but for now we simply note (for simplicity, we take
$2\times2$ $\hF$):
\beq
M(\hF\to\infty) \sim \lim_{F\to\infty}\left( {1\over F}+\ldots\right) \left(
1+F\gamma\gamma+\ldots\right) = \gamma\gamma
\eeq
This is in agreement with  eq.~(\ref{basic}):  the $\gamma\gamma$
factor here removes the corresponding $\gamma$-matrices in 
$\Gamma_{11}$ of (\ref{nineF}), reducing it to eq.~(\ref{basic}).

Eq.~(\ref{basic}) gives the supersymmetry projection imposed by D-branes
in a general constant background of NS fields.  (See~\cite{gg} for a
related discussion in light cone frame.) We now proceed to
construct BPS configurations of branes in such backgrounds. 

\section{Branes at Angles}
\label{sec:review}

In this section we review the general solution of branes at angles
with $\hF=0$, which was found in Ref.~\cite{bdl}.  Later we will
use T-duality on such configurations to produce others with $\hF\neq
0$ .  These situations will illustrate how conspiracies of the moduli
in Eq.~(\ref{basic}) can restore BPS saturation in configurations which
naively break supersymmetry.

To begin we establish some notation and conventions that are useful in
analyzing supersymmetry in the presence of D-branes.  Each D-brane
will induce a condition~(\ref{basic}). It is convenient to introduce a
Fock basis in which \beqn\label{eq:fock} a^j &=& {1\over2}(\Gm{2j} + i
\Gm{2j+1})\\ a^\dagger_j &=& {1\over2}(\Gm{2j} - i \Gm{2j+1}).  \eeqn
These operators satisfy the algebra $\{a^j,a_k^\dagger \} =\delta^j_k$
and so are lowering and raising operators. We denote the vacuum state
$|0\rangle$, which is annihilated by all $a^j$.  Note for later use
that $\Gm{2j}\Gm{2j+1}=-i(2n_j-1)$ where the $n_j$ are occupation numbers.
We will analyze the \SUSY projections induced by D-branes using the
raising and lowering operators introduced here.

\subsection{General Solution of  Branes at Angles}
\label{sec:soln}

Consider two $n$-branes\footnote{Following \cite{bdl}, $n$ refers
to the dimensions of the branes which do not overlap. Each brane
may or may not share additional dimensions.}
embedded in $2n$ compact dimensions where the two
branes are related by an $O(2n)$ rotation. We introduce complex
coordinates $\{z^i = x^{a_i} \pm i x^{b_i}\}$ where the $x$ are 
orthonormal real coordinates of the target space. Without loss
of generality, assume the first brane lies along ${\rm Re}\ z_i$.
Supersymmetry is preserved when the $O(2n)$ rotation acts as an $SU(n)$ 
or $ASU(n)$ rotation, for some choice of complex structure:
\beq
z^i \rightarrow R^k_l z^l
\eeq
In this case the two supersymmetry conditions reduce to:
\begin{equation}
\prod_{k=1}^n \left(a_k^\dagger + a_k\right) \tilde Q=\pm
\prod_k \left(R_k^{\dagger l} a_l^\dagger + R^k_l a^l \right) \tilde Q
\label{eq:cond}
\end{equation}
Solutions for the plus (minus) sign are $Q = \{|0\rangle, \: \prod_k
a_k^\dagger |0\rangle \}$ provided that $R$ is an $SU(n)$ ($ASU(n)$)
rotation.  Quantum numbers of the $a_k$ associated with the common
dimensions fill out the spinorial representations.

In particular, consider 10-dimensional models compactified on tori
$(T^2)^n$ (although this is a simplification, and we could easily turn
on moduli to get to $T^{2n}$). We will take an orthonormal frame with
$e_{2j}$ along the A-cycle of the $j^{th}$ $T^2$. We take the $a^{th}$
D-brane to lie along the directions
$\cos\theta_j^{(a)}e_{2j}+\sin\theta_j^{(a)}e_{2j+1}$.  Note that on
tori of finite size, there is a kind of quantization condition (or
rather a rationality condition) on the angles $\theta_j^{(a)}$: in
order not to be space-filling, the D-brane should lie along some
$(m_j,n_j)$-cycle of the tori. Thus it is simple to see that:
\begin{equation}
\tan\theta_j^{(a)}={m_j^{(a)}{\rm Im}\;\tau_j\over n_j^{(a)}+m_j^{(a)}
{\rm Re}\;\tau_j} 
\end{equation}
where $\tau_j$ is the modular parameter of the jth torus. The
supersymmetry conditions for each D-brane are then
\begin{equation} 
Q\pm \prod_{j=1}^n\left( \cos\theta_j^{(a)}\Gm{2j}
+\sin\theta_j^{(a)}\Gm{2j+1}\right)\tilde Q=0 
\end{equation} 
or 
\begin{equation}
Q\pm \prod_{j=1}^n\left( e^{-i\theta_j^{(a)}}a^{j} +e^{+i\theta_j^{(a)}}
a^\dagger_{j}\right)\tilde Q=0 
\end{equation}
We have assumed so far that each D-brane has undergone a $U(n)$
rotation from a reference configuration lined up along the A-cycles of
all tori. Ref.~\cite{bdl} found the condition for supersymmetry:
the rotation required to bring one D-brane in line with the other
should be in $SU(n)$ (or $ASU(n)$ in the case of opposite
orientation).  With the present conventions, for branes on $(T^2)^n$,
the $SU(n)$ case corresponds to the statement that
\begin{equation}
\sum_j (\theta_j^{(1)}-\theta_j^{(2)})=0\;\; {\rm
mod\ } 2\pi 
\label{sumtwo}
\end{equation}
for any pair of branes.  This is true up to changes in complex
structure: for example, if we complex conjugate one of the coordinates
(so that the corresponding angle changes sign) we still find
solutions, if the angles satisfy the corresponding condition. There is
also a suitable change in the Fock states.

\section{T-duality}
\label{sec:examples}
In this section we discuss the action of T-duality on the BPS
configurations of branes at angles from Sec.~\ref{sec:soln}.   We will
see that the more general formulation of eq.~(\ref{basic}) is necessary
to consistently account for BPS saturation after T-duality.

\subsection{2-2 versus 0-4}
\label{sec:22angle}
Our first example concerns a BPS saturated configuration of 2-branes at
angles.  T-duality performed along directions spanned by one of the
branes turns the 2-2 configuration into a 0-brane bound to a 4-brane
with gauge fields on the 4-brane.  The strength of the gauge fields is
directly related to the angles in the 2-2 configuration. The 0-4
system is BPS by itself~\cite{Dnotes} and so we must show that gauge
fields produced by T-duality do not spoil the supersymmetry due to
their appearance in the \SUSY relation~(\ref{basic}).

\subsubsection{2-2}
\label{sec:22}
We consider two 2-branes on a $(T^2)^2$ in the 6789-directions, in
the setup described in Sec.~\ref{sec:soln}. The supersymmetry
conditions reduce to:
\begin{equation}
\tilde Q=\pm
e^{-i\sum_{j=3}^4
[\theta_{j}^{(2)} - \theta_j^{(1)}]
(2n_j-1)}\;\tilde Q
\label{eq:thetasoln}
\end{equation}
Thus in the case that the branes have the same orientation, 
we have solutions when:
\begin{equation}
\sum_{j=3}^4 (\theta_j^{(1)} - \theta_j^{(2)}) (2n_j - 1) = 0 \bmod 2\pi
\label{eq:22soln}
\end{equation}
For $n_3 = n_4$ this yields the same solutions as eq.~(\ref{sumtwo}) and
the surviving spinors are $\tilde Q=|00\rangle, |11\rangle =
a_3^\dagger a_4^\dagger |00\rangle$.  For $n_3 \neq n_4$ the surviving
spinors are $|10\rangle$ and $|01\rangle$.  
The latter solution corresponds to a change of complex structure,
in which the complex coordinate of
one of the $T^2$ factors is conjugated.
In the case of opposite orientation, we have solutions when
\begin{equation}
\sum_{j=3}^4 (\theta_j^{(1)}-\theta_j^{(2)}) (2n_j -1)=\pi\;\;  \bmod 2\pi
\label{sumtwob}
\end{equation}
which is the modification of eq.~(\ref{sumtwo}) and eq.~(\ref{eq:22soln})
for an $ASU(n)$ rotation.  This result has a simple interpretation: a
brane-antibrane pair at angles is identical to a brane-brane
configuration at different angles: the solution in eq.~(\ref{sumtwob})
is obtained by turning a brane into an antibrane by a $\pi$-rotation 
in one 2-plane. 

\subsubsection{0-4}
\label{sec:04}
The 2-2 configuration of the previous section can be turned by
T-duality into a 0-brane bound to a 4-brane with gauge fields on the
4-brane.  In this section we study possible solutions to the
supersymmetry condition~(\ref{basic}) for the 0-4 case.  For simplicity,
consider a 4-brane wrapped on a $T^4$ in the 6789 directions  with a gauge
field of the form:
\begin{equation}
\hF=\pmatrix{f_3\sigma&0\cr 0& f_4\sigma} 
\: \: \: \: \: \: \: \: \: \:
\sigma = \pmatrix{0 & 1 \cr -1 & 0}
\end{equation}
$\hat F$ modifies the supersymmetry projections imposed by the 4-brane and
also induces 2-brane and 0-brane charge via the 
Chern-Simons couplings on the 4-brane.
Given the general formulation of Section~\ref{sec:setup}, after a
little algebra the supersymmetry conditions read:
\begin{eqnarray}
Q\pm\tilde{Q}=0\\
Q\pm \prod_{j=3}^4 (-i)(2n_j-1)\sqrt{{1+if_j(2n_j-1)\over
1-if_j(2n_j-1)}} 
\;\tilde{Q}=0
\end{eqnarray}
If we define $f_j=-\cot\alpha_j$, the second equation
becomes:
\begin{equation}
Q\pm e^{i\sum_{j=3}^4 \alpha_j(2n_j-1)}\;\tilde Q=0
\end{equation}
(Note here that $0 < \alpha_j < \pi$ corresponds to finite
$f_j$.)  Thus, in the brane-brane (brane-antibrane) case, the two
conditions are compatible if: 
\begin{equation}
\sum_{j=3}^4 \alpha_j(2n_j-1)=0 \; (\pi) \bmod 2\pi\
\label{eq:44soln}
\end{equation}
This should be compared to Eqs.~(\ref{eq:22soln}) and~(\ref{sumtwob}).
Because of the restriction in the range of $\alpha_j$, we conclude
that the brane-antibrane (BA) case is solved by the spinors
$|00\rangle$ and $|11\rangle$ for which $\alpha_3 + \alpha_4 = \pi
\bmod 2\pi$.  In contrast, the brane-brane (BB) case has solutions
$|01\rangle$ and $|10\rangle$ for which $\alpha_3 - \alpha_4 = 0 \bmod
2\pi$.  In terms of the physical gauge field $F$, the BA and BB cases
correspond to anti-self-dual ($f_3 = -f_4$) and self-dual ($f_3 =
f_4$) fields.  This is related to the fact that 0-branes and
anti-0-branes marginally bound to 4-branes can be understood as small
instantons and anti-instantons of the 4-brane gauge
theory~\cite{douglas95a}.  We have seen in Sec.~\ref{sec:setup} that
in the presence of a constant gauge field background the
supersymmetries that are present on the brane are different from the
\SUSYs that survive in the absence of the background.  The BPS
configurations derived in this section are self-dual in order that the
\SUSYs that survive the introduction of the 0-brane (instanton) on the
4-brane are a subset of the \SUSYs that are present on the 4-brane with
constant gauge fields.  We can also understand this from T-duality.
Whether T-duality of a 2-2 configuration produces a 0-brane or an
anti-0-brane bound to a 4-brane depends on the relative orientations
of the original 2-branes after the $SU(2)$ ($ASU(2)$) rotations.  This is
also reflected in the gauge fields produced by T-duality.

\subsubsection{T-duality}
\label{sec:22dual}
Now we show that T-duality maps the 2-2 configurations of Sec.~\ref{sec:22}
into precisely the BPS 0-4 configurations with gauge fields described
above.  Indeed, the angles between branes in the 2-2 case are mapped
directly into gauge fields on the 4-brane.  It is easiest to study
T-duality by examining its effects on the boundary conditions for  open
strings propagating on the D-branes. (With a little more effort it is
possible to examine the effects of T-duality directly on the
background metric of the bent torus and the antisymmetric tensor.
That analysis is consistent with the vertex operator manipulations
below, and will not be presented here.)  

Let us perform T-duality along the dimensions spanned by one of the
two branes in Sec.~\ref{sec:22}.   Without loss of generality we can
pick orthonormal coordinates $\{X^6,\cdots X^9\}$ where the first
brane lies along $X^6$ and $X^8$.  Then open strings propagating on
the two branes have boundary conditions:
\begin{eqnarray}
1:& \partial_n X^{6,8}  =  \partial_\tau X^{7,9}  = 0 \nonumber \\
2:& 
\partial_n \left[\cos\alpha_{3,4} \; X^{6,8} + 
            \sin\alpha_{3,4} \; X^{7,9}\right]  = 0 \nonumber \\
\: & \partial_\tau \left[-\sin\alpha_{3,4} \; X^{6,8} + 
            \cos\alpha_{3,4} \; X^{7,9}\right]  = 0 
\end{eqnarray}
where we have defined $\alpha_{3,4} = \theta_{3,4}^{(2)} -
\theta_{3,4}^{(1)}$.   T-duality exchanges Neumann and Dirichlet
boundary conditions.  So, T-dualizing along $X^6$ and $X^8$ gives:
\begin{eqnarray}
1:& \partial_\tau X^{6,7,8,9}  = 0 \nonumber \\
2:& 
\partial_n X^{7,9} + \cot\alpha_{3,4}\;  \partial_\tau X^{6,8} = 0
\nonumber \\
\:  &\partial_n X^{6,8} - \cot\alpha_{3,4}\;  \partial_\tau X^{7,9} = 0
\end{eqnarray}
These boundary conditions can be interpreted as a 0-brane or an anti
0-brane bound to a 4-brane with a gauge field $F_{76} = \cot\alpha_3$
and $F_{98} = \cot\alpha_4$~\cite{rgl}.  The two possible BPS
conditions for 2-branes at angles $\alpha_3 \pm \alpha_4 = 0$
translate into precisely the self-dual and anti-self-dual solutions
in the 4-4 case.

\subsection{3-3 versus 1-5}
In the example in the previous section T-duality of the 2-branes
produced a 0-brane bound to a 4-brane.  This configuration is already
BPS saturated and our task was to show that the resulting gauge fields
did not interfere with BPS saturation via their appearance in the
supersymmetry projection Eq.~(\ref{basic}).  An even more instructive
example arises from T-duality of 3-branes at angles.  These can be
converted to a 1-brane at an angle relative to a 5-brane.   Such a
configuration would normally break all the supersymmetries.  We show
here that the gauge fields that are simultaneously produced by
T-duality restore BPS saturation via action of the matrix $M(F)$
appearing in  the supersymmetry relation~(\ref{basic}).

\subsubsection{3-3}
We consider two 3-branes\footnote{Again, these branes may intersect
in additional non-compact dimensions.} 
on a $(T^2)^3$ in the 456789-directions, in
the setup described in Sec.~\ref{sec:soln}. The supersymmetry
conditions reduce to:
\begin{equation}
\tilde Q=\pm
e^{-i\sum_{j=2}^4
[\theta_{j}^{(2)} - \theta_j^{(1)}]
(2n_j-1)}\;\tilde Q
\label{eq:33cond}
\end{equation}
When the branes have the same orientation, we have solutions when:
\begin{equation}
\sum_{j=2}^4 (\theta_j^{(1)} - \theta_j^{(2)}) (2n_j - 1) = 0 \bmod 2\pi
\label{eq:33soln}
\end{equation}
For $n_2 = n_3 = n_4$ we find the same solutions as in Eq.~(\ref{sumtwo}) 
and the surviving spinors are $\tilde{Q} = |000\rangle,
\, |111\rangle = a_2^\dagger a_3^\dagger a_4^\dagger |000\rangle $.
Filling out the spinorial representations in the four non-compact
dimensions yields $N=1$ \SUSY in $d=4$ for the generic case where the
3-branes intersect at a point.  The solutions for unequal $n_i$
preserve the same amount of supersymmetry albeit with different
surviving spinors and correspond, as in the 2-2 case, to SU(3)
rotations in complex structures where the complex coordinates of some
of the $T^2$ factors have been conjugated.  Again as in the 2-2 case,
the solutions in the case of opposite orientations replace the right
side of Eq.~(\ref{eq:33soln}) with $\pi \bmod 2\pi$ and reflect the fact
that a brane is turned into an anti-brane by a rotation by $\pi$ in
one 2-plane.

An interesting phenomenon in the 3-3 case that differs from the 2-2
situation is \SUSY enhancement in some parts of the moduli space of BPS
configurations.  For example, if $\theta_2^{(2)} - \theta_2^{(1)} = 0$,
so that the 3-branes intersect along a line rather than on a point,
the supersymmetry in enhanced to $N=2$ in $d=4$ because the spinors
with $n_3 = n_4$ or $n_3 \neq n_4$ in the above analysis will have the
solutions Eq.~(\ref{eq:33soln}) regardless of the value of $n_2$.

\subsubsection{1-5}
\label{sec:15}
T-duality of 3-3 configurations of the previous section along two of
the dimensions spanned by one of the D-branes yields a 1-brane at an
angle to a 5-brane with a gauge field background.  Consider a 5-brane
on a $T^5$ in the 56789 directions with gauge fields of the form:
\begin{equation}
\hF=\pmatrix{0 & 0 & 0 \cr 
             0 & f_3\sigma & 0 \cr 
             0 & 0 & f_4\sigma } 
\: \: \: \: \: \: \: \: \: \:
\sigma = \pmatrix{0 & 1 \cr -1 & 0}
\end{equation}
Besides modifying the supersymmetry projections imposed by the 5-brane,
$\hat F$ induces 3-brane and 1-brane charge.
Now add a 1-brane at an angle $\alpha_2$ relative to the 5-brane on
the $T^2$ in the 45-directions.  Then the general formulation of
Sec.~\ref{sec:setup} and an analysis parallel to Sec.~\ref{sec:04}
yields the supersymmetry condition:
\begin{equation}
\tilde{Q} = \pm e^{i \sum_{j=2}^{4} \alpha_j (2n_j -1)} \tilde{Q}
\end{equation}
where we have defined $f_j = -\cot\alpha_j$ for $j=3,4$ so that $0 <
\alpha_j < \pi$ as in Sec.~\ref{sec:04}.  Therefore solutions exist
in the brane-brane (brane-antibrane) case when
\begin{equation}
\sum_{j=2}^4 \alpha_j (2n_j - 1) = 0 (\pi) \bmod 2\pi
\label{eq:15soln}
\end{equation}
which should be compared to eq.~(\ref{eq:33soln}).  If we take
take $\alpha_2 \geq 0$,  then given the restriction of the range
of $\alpha_3$ and $\alpha_4$, the brane-antibrane (BA) case is
solved by the spinors $|000\rangle$ and $|111\rangle$ when $\alpha_2 +
\alpha_3 + \alpha_4 = \pi \bmod 2\pi$.\footnote{Additional solutions exist
which reverse the signs of the $\alpha_j$ with corresponding changes
in the spinors.}  The BB case is solved by the spinors $|110\rangle$
and $|001\rangle$ when $\alpha_2 + \alpha_3 - \alpha_4 = 0$.   These
are solutions with $N=1$ supersymmetry in the four non-compact dimensions.

It is instructive to examine the limit $\alpha_2 \rightarrow 0$ in
which the 1-brane lies within the 5-brane.   In this case, the
restrictions on the range of $\alpha_3$ and $\alpha_4$ (which are
equivalent to requiring finite field strengths on the 5-brane) gives
solutions in the BB case when $\alpha_3 - \alpha_4 = 0$ and in the BA
case when $\alpha_3 + \alpha_4 = \pi$.  In terms of the physical gauge
fields these correspond to self-dual $f_3 = f_4$ and anti-self-dual
$f_3 = -f_4$ gauge fields in the four dimensions on the 5-brane that
are transverse to the 1-brane.  This self-duality arises for the same
reason as the self-duality of the gauge fields in the 0-4
configurations in Sec.~\ref{sec:04}. Furthermore, the self-dual and
anti-self-dual fields are consistent with four choices of spinors
each: in the BB case solutions are $|110\rangle, \, |010\rangle, \,
|101\rangle$ and $|001\rangle$.   After filling out the spinorial
representations in the non-compact dimensions this gives $N=2$ \SUSY in
$d=4$.  In other words, as the relative angle between the 1-brane and
5-brane is decreased to zero we are seeing an enhancement of
supersymmetry.\footnote{We thank M.~Cveti\v{c} for discussions
regarding this point and its analogue in the language of classical p-brane
solutions.}  

We know from Ref.~\cite{douglas95a} and from the Chern-Simons terms in
the D-brane effective action that a 1-brane within a 5-brane can
be understood as self-dual gauge field in the four dimensions
transverse to the 1-brane on the 5-brane worldvolume.   
The conditions $\alpha_2 - \arccot f_3 + \arccot f_4 = 0$ and $\alpha_2
-\arccot f_3 - \arccot f_4 = \pi$ that solve the BB and BA cases can be
understood as a generalization of self-duality.   Branes at angles
arise in worldvolume terms as gauge fields shifted away from
self-duality in a specific way.

\subsubsection{T-duality}
\label{sec:33T}
To show the T-dual relation between the 3-3 and 1-5 configurations
displayed above we repeat the analysis of Sec.~\ref{sec:22dual} that
related the 2-2 and 0-4 configurations.  We only want to T-dualize
along two of the dimensions of one of the 3-branes in order to turn it
into a 1-brane.  So the treatment is identical to Sec.~\ref{sec:22dual}
and the 3-3 system T-dualizes to a 1-5 system with $F_{76} = \cot
\alpha_3$ and $F_{98} = \cot \alpha_4$.  Here $\alpha_{3,4} =
\theta_{3,4}^{(2)} - \theta_{3,4}^{(1)}$.  This exactly matches the
1-5 configurations of the previous section.

\section{The Field Theory Perspective}
\label{sec:gauge}
In this section we re-examine the question of BPS saturation from the
point of view of the world-brane effective theory.  The world-brane
theory has the field content of the dimensional reduction of $N=1$
super Yang-Mills from 10 dimensions down to the brane~\cite{witt1}.
To check if supersymmetry is preserved in the presence of a
nonvanishing $F$ background we must study the variation of the
gaugino.  The super Yang-Mills variation of the gaugino is given by
\begin{equation}
\delta_\epsilon \chi = (-1/4) \Gamma^{\mu\nu} F_{\mu\nu} \epsilon.
\label{eq:gaugvar}
\end{equation}
This would seem to be in direct contradiction of
the assertion in Eq.~(\ref{basic}) that {\em any} constant gauge field
(or NS 2-form) background on a single D-brane is a BPS state that merely
redefines the \SUSYs surviving the presence of the brane.  
In this section we resolve this conundrum by showing the the physical
gaugino vertex operator is modified in the presence of background
gauge fields and that the variation of this field under the surviving
supersymmetries vanishes identically. (Related work appears in
Ref. \cite{gg}.)

\subsection{Light Cone Gauge}
\label{sec:lightcone}
It is easiest to carry out the analysis in light-cone gauge with 10
dimensional fields.  The analysis for any D-brane can be carried out
by a straightforward dimensional reduction of our fields to the brane
world-volume.  To begin we recall the light-cone decomposition of
fields and supersymmetries that appears in Sec. 5.3 of ~\cite{gsw}.
The left and right moving massless boson vertex operators transform in
the $8_v$ representation of the transverse SO(8) in light-cone frame
and are denoted $V_{BL}^i$ and $V_{BR}^i$, where $i=1\cdots 8$.
States are constructed by the action of $V_{B}^i \, \zeta^i$ where $\zeta^i$
is the boson wavefunction.  Ten dimensional 16-component Majorana-Weyl
spinors decompose under the transverse SO(8) into 8-component dotted
and undotted spinors that tranform under the $8_s$ and $8_c$
representations.  So we can split the left and right moving
supersymmetries into $Q^a$, $Q^{\dot{a}}$, $\tilde{Q}^a$ and
$\tilde{Q}^{\dot{a}}$.  The massless spinor (gaugino) vertex operator
can be split into dotted and undotted pieces, but the Dirac equation
implies that the undotted fermion wavefunction is fully determined in
terms of the dotted fermion components.  So, on shell, fermion states
are created by left and right moving operators $V_{FL}^{\dot{a}}
\, u^{\dot{a}}$ and $V_{FR}^{\dot{a}} \, u^{\dot{a}}$ where $u^{\dot{a}}$ is
the gaugino wavefunction.

The action of the undotted supersymmetry on the  bosons is summarized
by~\cite{gsw,gg}: 
\begin{eqnarray}
\left[\eta^a \, Q^a, \, V_{BL}^i\, \zeta^i \right] = &
V_{FL}^{\dot{a}} \, \tilde{u}^{\dot{a}} 
 \nonumber \\
\left[\eta^a \, \tilde{Q}^a, \, V_{BR}^i\, \zeta^i \right] = &
V_{FR}^{\dot{a}} \, \tilde{u}^{\dot{a}} 
\label{eq:susyact}
\end{eqnarray}
where $\tilde{u}^{\dot{a}}(\eta^a, \zeta^i) = k^+ \: \eta^a
\,\gamma^i_{a\dot{a}}\, \zeta^{i}$.  Here $k^i$ and $k^+$ are transverse
and light-cone components of the momentum in the boson and the gamma
matrices are in the representation given in Appendix 5.B
of~\cite{gsw}.  (The dotted \SUSYs $Q^{\dot{a}}$ act in a similar
fashion except that $\tilde{u}$ is different in this case.)  The
transformed fermion wavefunction $\tilde{u}$ gives precisely the light
cone decomposition of the ten dimensional super Yang-Mills gaugino
variation in Eq.~\ref{eq:gaugvar}.
In what follows we will construct the physical gaugino operator on a
D-brane with background fields and show that its commutator with the
surviving supersymmetry in Eq.~(\ref{basic}) vanishes.

\subsection{Construction of Vertex Operators}
\label{sec:vertex}
In the presence of a gauge field background, the bosonic
coordinate $X^i = X_L^i(\tau - \sigma) + X_R^i(\tau + \sigma)$ of open
strings propagating on the brane satisfying a boundary condition of
the form $X^i_L = M_{ij} X_R^j$ where $M$ is the light-cone vector
representation of the rotation $-\Omega(\Gamma)(1 - F)/(1 + F)$ that
appears in eq.~(\ref{basic})~\cite{gg}.    Define a new coordinate
$\bar{X}_L^i = M_{ij} X_L^j$.     Then the left and right moving parts
of the dual coordinate $\bar{X}^i = \bar{X}_L^i + X_R^i$ satisfy
Neumann boundary conditions.   The bosonic part of the vertex operator
for the massless 
boson is therefore proportional to $\partial_\tau \bar{X}^i =
\partial_\tau \bar{X}_L^i + \partial_\tau X_R^i$.   In terms of the
original coordinate $X$, this means that the operator in
proportional to $ [ (\partial_\tau + \partial_\sigma) X^i +
M^{-1}_{ij} (\partial_\tau - \partial_\sigma) X^j]$.   Multiplying by
an overall factor of $M$, and supersymmetrizing the vertex operator
tells us that the massless boson vertex operator that satisfies the
boundary conditions imposed by the gauge field is proportional to the 
the following combination left and right moving operators:
\begin{equation}
V_B^i = V_{BL}^i + M_{ij} V_{BR}^j
\end{equation}
Supersymmetry then gives us a similar expression for the physical
gaugino vertex operator:
\begin{equation}
V_F^{\dot{a}} = V_{FL}^{\dot{a}} + M_{\dot{a} \dot{b}} V_{FR}^{\dot{b}}
\end{equation}
where $M_{\dot{a}\dot{b}}$ is an SO(8) dotted spinor representation of the
rotation $M$, the lightcone decomposition of eq. (\ref{M}).  
Note that we can construct another combination of
$V_{FL}$ and $V_{FR}$:
\begin{equation}
\phi_F^{\dot{a}} = V_{FL}^{\dot{a}} - M_{\dot{a} \dot{b}} V_{FR}^{\dot{b}}
\end{equation}
but $\phi$ vanishes when acting on a physical state because of
boundary conditions imposed by the presence of the gauge field.
Finally, the light cone decomposition of the supersymmetry condition
in eq.~(\ref{basic}) is:
\begin{equation}
Q_{+}^a |B\rangle = (Q^a - M_{ab} \tilde{Q}^b ) |B\rangle = 0
\end{equation}
with a similar equation for the dotted \SUSYs~\cite{gg}.

\subsection{Variation of the Gaugino}
As discussed in Sec.~\ref{sec:lightcone}, in order to study the 
variation of the gaugino,  we commute the supercharge with the massless
boson vertex operator, and read off the wavefunction of the resulting
fermion.  Using Eq.~(\ref{eq:susyact}) for the action of supersymmetries
we find that:
\begin{eqnarray}
\left[ \eta^a \, Q^a_+, V_B^i \, \zeta^i \right] &=&
V_{FL}^{\dot{a}} \, \tilde{u}^{\dot{a}}(\eta^a,\zeta^i) -
V_{FR}^{\dot{a}} \, \tilde{s}^{\dot{a}}(\eta^a,\zeta^i) 
\\ 
\tilde{s}^{\dot{a}}(\eta^a,\zeta^i) &=&
\tilde{u}^{\dot{a}}(\eta^c \, M_{ca}, \, \zeta^k \, M_{ki})
\end{eqnarray}
where $\tilde{u}^{\dot{a}}(\eta^a,\zeta^i)$ is defined below
eq.~(\ref{eq:susyact}).  Using the definition of $V_F$ and $\phi_F$ in
terms of $V_{FL}$ and $V_{FR}$ we can rewrite this as:
\begin{equation}
\left[ \eta^a Q^a_+, V_B^i \zeta^i \right] = (1/2) V_F^{\dot{a}}
\left[ \tilde{u}^{\dot{a}} - M_{\dot{a}\dot{b}} \tilde{s}^{\dot{b}}
\right]
\equiv
(1/2) V_F^{\dot{a}} t^{\dot{a}}
\end{equation}
where we have dropped terms proportional to $\phi_F$ since this
operator evaluates to zero on physical states.  The wavefunction
$t^{\dot{a}}$ can be decomposed as:
\begin{equation}
t^{\dot{a}} = k^+ 
\eta^a
\left[ \gamma_{a\dot{a}}^i - M_{ij} M_{\dot{a}\dot{b}} M_{ab}
\gamma^j_{b\dot{b}} \right]
\zeta^i
\end{equation}
The quantity in parentheses relates~\cite{gg} the vector, dotted
spinor and undotted spinor representations of SO(8) rotations and
vanishes {\em identically}. As discussed in Sec.~\ref{sec:lightcone},
the wavefunction $t^{\dot{a}}$ is the lightcone decomposition of
$\delta\chi$, the variation of the physical gaugino field.  We have
therefore shown that the variation of the physical gaugino under the
unbroken undotted supersymmetries vanishes identically.  The dotted
\SUSYs can be treated similarly.

The physical significance of this is clear: it is not enough, in
seeking to identify all supersymmetric configurations, to simply look
at the supersymmetry variations of super-Yang-Mills.  The background
field induces important mixing effects, both in the generator of
supersymmetry, as well as the physical mass eigenstates.  Presumably,
in the presence of a constant background field, it is necessary to
consider a supersymmetrized version of the full Dirac-Born-Infeld
action which is correct to all orders in $\alpha'$.

\section{Classical Solutions and Black Hole Entropy}
\label{sec:classic}
In this paper we have demonstrated that there are compactifications of
D-branes at angles that are supersymmetric only when suitable
antisymmetric tensor moduli fields are turned on at the position of the
brane.  It would be very attractive to demonstrate this from
the point of view of the classical p-brane solutions of supergravity.
To date, the general procedure for constructing intersecting brane
solutions is only understood for orthogonal branes~\cite{ggetc} and
for branes embedded in larger branes~\cite{embed}.  Furthermore, there
are no available intersecting p-brane solutions where the branes are
localized on each other.  Both of these difficulties impede an
analysis of the issues of this paper from the viewpoint of classical
solutions.  The p-brane solutions corresponding to the D-brane
configurations presented here will have the unusual property that the
antisymmetric tensor field moduli will have to converge to some
specific value as the core of the solution is approached regardless of
the asymptotic value.  When the intersecting p-branes at angles are
compactified, this should yield a class of ``fixed scalar'' fields in
the non-compact space.  (See~\cite{fixed} for discussions of such
``fixed scalars'' and their appearance in the physics of black holes.)

The results of this paper also have interesting applications to 
computations of extremal black hole entropy in Type II string theory.
The generating solution for NS-NS charged black hole solutions of Type
II on $T^6$ has been discussed in~\cite{ct,ch}, and 
contains five quantized charges as opposed to the four that have
appeared in the extant computations of four dimensional black hole
entropy~\cite{bhetc}.  The fifth charge, which is necessary in order
to account for the $E_{7(7)}$ symmetry of the complete entropy formula, is
associated with branes at angles.  Indeed, the solutions
of~\cite{ct,ch} contain fundamental strings at angles relative to NS
5-branes and various dualities can be applied to produce black holes
containing D-branes at angles.  Our work has shown how configurations
of branes at angles can be made supersymmetric by turning on NS 2-form
backgrounds. These backgrounds also induce RR-charges via Chern-Simons
couplings
on the brane world-volume. These charges must be accounted for in the
$E_{7(7)}$ counting of states. The dualities that produce the
D-branes also produce Ramond-Ramond background fields and so it is
necessary to understand these also in order to account for the black hole
entropy.  Work is in progress to understand these issues as well as to
construct the classical solutions discussed in the previous
paragraph~\cite{vbrl}.

\section{Conclusion}
In this paper we have shown that non-vanishing antisymmetric tensor
backgrounds modify the supersymmetry projections imposed by D-branes.
These modifications are necessary for the consistency of T-duality 
since BPS configurations of branes at angles are dual to
branes with antisymmetric tensor backgrounds.  We argued that the
realization of supersymmetry on the brane must also be modified in the
presence of an NS 2-form and showed, in particular, that the
variation of the physical gaugino vanishes in any constant 
background.   One consequence of these results is the
existence of unusual BPS configurations of branes at angles - for
example, a 1-brane at an angle to a 5-brane is supersymmetric if a
suitable background is turned on.   The 1-5 system is also interesting
in that the generic configuration has N=1 supersymmetry in the four
non-compact dimensions, and these configurations are related to
N=2  systems where the 1-brane lies within the 5-brane.
We also sketched a forthcoming application of this work to the study
of black hole entropy in string theory~\cite{vbrl}.

\acknowledgments
We are grateful to
C. Callan, M. Cveti\v{c}, M. Douglas, G. Lifschytz, S. Ramgoolam and
H. Verlinde for helpful discussions.  We also wish to thank the
organizers of {\em Strings '96} where this work was begun as well as
the Aspen Center for Physics.  V.B. is supported in part by DOE grant
DE-FG02-91-ER40671. R.G.L. was  supported in part by DOE grant
DE-FG05-90ER40559.

\end{document}